# Testing the chondrule-rich accretion model for planetary embryos using calcium isotopes

Elsa Amsellem[1], Frédéric Moynier[1,2], Emily A. Pringle[1], Audrey Bouvier[3], Heng Chen[4], James M. D. Day[1,5]

[1]Institut de Physique du Globe de Paris, Université Paris Diderot, Sorbonne Paris Cité, CNRS UMR 7154, 1 rue Jussieu, 75238 Paris, France

[2]Institut Universitaire de France, Paris, France.

[3]Department of Earth Sciences, Centre for Planetary Science & Exploration, University of Western Ontario, London, ON, N6A 3K7 Canada

[4]Department of Earth and Planetary Science, Washington University in St Louis, St Louis, MO 63130, USA

[5]Scripps Institution of Oceanography, University of California San Diego, La Jolla, CA 92093-0244, USA

Corresponding author: Elsa Amsellem (amsellem@ipgp.fr)

# Abstract

Understanding the composition of raw materials that formed the Earth is a crucial step towards understanding the formation of terrestrial planets and their bulk composition. Calcium is the fifth most abundant element in terrestrial planets and, therefore, is a key element with which to trace planetary composition. However, in order to use Ca isotopes as a tracer of Earth's accretion history, it is first necessary to understand the isotopic behavior of Ca during the earliest stages of planetary formation.

Chondrites are some of the oldest materials of the Solar System, and the study of their isotopic composition enables understanding of how and in what conditions the Solar System formed. Here we present Ca isotope data for a suite of bulk chondrites as well as Allende (CV) chondrules. We show that most groups of carbonaceous chondrites (CV, CI, CR and CM) are significantly enriched in the lighter Ca isotopes ($\delta^{44/40}Ca$ = +0.1 to +0.93‰) compared with bulk silicate Earth ($\delta^{44/40}Ca$ = +1.05 ± 0.04‰, Huang et al., 2010) or Mars, while enstatite chondrites are indistinguishable from Earth in Ca isotope composition ($\delta^{44/40}Ca$ = +0.91 to +1.06‰). Chondrules from Allende are enriched in the heavier isotopes of Ca compared to the bulk and the matrix of the meteorite ($\delta^{44/40}Ca$ = +1.00 to +1.21‰). This implies that Earth and Mars have Ca isotope compositions that are distinct from most carbonaceous chondrites but that may be like chondrules. This Ca isotopic similarity between Earth, Mars, and chondrules is permissive of recent dynamical models of planetary formation that propose a chondrule-rich accretion model for planetary embryos.

# 1. Introduction

Chemical and isotopic studies of meteorites are a primary source of information for understanding Solar System formation and evolution, from the nebula phase to the formation of planets. Many meteorites represent materials from the earliest stages of planetary formation and thus their study informs on the origin of terrestrial planets (e.g. Moynier and Fegley Jr, 2015), even if the present-day inventory of meteorites may not necessarily be representative of the accreting materials swept into the planets to form them. The relative chemical abundance of refractory elements in chondrites are close to that of the Sun's photosphere composition and therefore also the solar nebula, making chondrites a key source of information about the material that accreted to form the planets (e.g. Palme et al., 2014). Isotopic studies, particularly those using oxygen isotopes, have enabled significant advancement in understanding the relationship between Earth, the Moon and chondrites. Nevertheless, the composition of the bodies that formed the Earth, and the chemical relationship between the Earth and the Moon are still subject to debate (e.g. Dauphas et al., 2014).

Among major elements constituting inner Solar System material, only O and Ca exhibit significant isotopic fractionation among chondrite groups and between chondrites and the bulk silicate earth (BSE) (Clayton and Mayeda, 1999; Simon and DePaolo, 2010; Huang and Jacobsen, 2012; Valdes et al., 2014). Silicon isotopes are mostly fractionated between terrestrial rocks and chondrites (Georg et al., 2007; Savage and Moynier, 2013) and in highly volatile-depleted differentiated meteorites (e.g. Pringle et al., 2014), while other major elements, such as Mg, do not show significant fractionation between chondrites and the BSE (e.g. Teng et al., 2010; Bouvier et al., 2013).

Calcium has five stable isotopes, $^{40}$Ca (96.94%), $^{42}$Ca (0.647%), $^{43}$Ca (0.135%), $^{44}$Ca (2.086%) and $^{46}$Ca (0.004%), and one long-lived radioactive isotope that has such a long half-life that it is essentially stable ($^{48}$Ca; 0.187%; half-life = 4.3×10$^{19}$ years). Calcium, despite its relatively high mass, shows large isotopic fractionation up to 1‰ for the $^{44}$Ca/$^{40}$Ca ratio in the BSE, the Moon, Mars, chondrites and differentiated meteorites (Russell et al., 1978; Niederer and Papanastassiou, 1984; DePaolo, 2004; Simon and DePaolo, 2010; Huang et al., 2012; Valdes et al., 2014; Magna et al., 2015). As Ca is both a lithophile and refractory element (50% condensation T = 1659K, Lodders et al., 2003), it does not partition into the core, nor it is volatilized during planetary accretion. Therefore its isotopic composition is *a priori* not modified during planetary formation and differentiation of the BSE. These characteristics coupled with the recent development of routine high precision Ca isotopic measurement have led to its use in understanding the composition and origins of the terrestrial planets (Simon and DePaolo, 2010; Huang and Jacobsen, 2012; Valdes et al., 2014). However, high temperature fractionation has been reported for mantle peridotites and co-existing minerals, which may account for some of the variation found in igneous rocks (Huang and Jacobsen, 2012).

Previous Ca isotopic studies showed that different carbonaceous chondrite groups are isotopically distinct and that all groups except the CO chondrites are enriched in the lightest isotopes of Ca, when compared with Earth (Simon and DePaolo, 2010; Huang et al., 2012; Valdes et al., 2014). The origin of the isotopic variations between carbonaceous chondrite groups is puzzling. Although most Ca-Al-rich inclusions (CAIs) are enriched in the lightest isotopes of Ca compared to Earth (Niederer and Papanastassiou, 1984; Huang et al., 2012), the isotopic compositions of bulk carbonaceous chondrites do not correlate with CAI abundances. For example, CI chondrites do not contain CAIs, but are isotopically lighter than CO chondrites, which contain ~10% by volume of CAIs (Zanda et al., 2006). Chondrules are the main carriers of Ca in chondrites (>45% of the total Ca, Rubin, 2011) but, to date, only one high-precision Ca isotope value has been reported for an individual chondrule (from the Allende chondrite) (Simon et al., 2016). This single datum shows that chondrules may be enriched in the heavier isotopes of Ca when compared to the bulk, by about 0.5‰ for the $^{44}Ca/^{40}Ca$ ratio.

Recent models have proposed that planetary embryos formed in chondrule-rich regions and that the direct accretion of chondrules would be a viable mechanism to enable rapid growth of planetary embryos up to the size of Mars in the terrestrial planet formation region (Johansen et al., 2015). Since Ca isotopes are not fractionated during planetary accretion, but may be fractionated between matrix, CAIs and chondrules, they can be used to test the model of chondrule accretion. This model predicts that ~90% of the mass of the Earth could be inherited from planetary embryos formed by chondrule accretion (Johansen et al., 2015); therefore, the Ca isotopic composition of Mars and the Earth should be close to the composition of chondrules.

In this study, we present high precision Ca isotopic composition measurements of a variety of terrestrial basalts, bulk chondrites, three individual chondrules and a CAI by multiple-collector inductively-coupled-plasma mass-spectrometry (MC-ICPMS), using the standard bracketing technique, in order to understand the origin of the variations among meteorites and to test the chondrule accretion model for planetary embryos.

## 2. Materials and Methods

*2.1 Samples*

Samples were chosen to represent a wide range of carbonaceous chondrites: Allende (CV3), Asuka 881595 (CR2), Cold Bokkeveld (CM2), Yamato 980115 (CI1), Orgueil (CI1), Pecora Escarpment (PCA) 02012 (CM2) and PCA 02010 (CM2). Two enstatite chondrites were measured: Indarch (EH4) and Khairpur (EL6). We also measured a mid-ocean ridge basalt (MORB) from the South Mid-Atlantic Ridge (EW9309 10D), two ocean island basalts (OIB) from the Canary Islands (Spain) (Day et al., 2010), one OIB from Hawaii, USA (BHVO-2) and a continental flood basalt from the Colombia River, USA (BCR-2). We also measured a refractory harzburgite xenolith from Lanzarote in the Islands (LZ0604B) with very low loss on ignition (0.04 wt.%) and low $Al_2O_3$ (0.76 wt.%), and a partially serpentinized harzburgite standard from the USGS (PCC).

In addition to bulk rock samples, one individual CAI from Allende called AB1 was analysed. It is a fine-grained CAI with a type III REE pattern and has been used for a previous isotopic study of Ba (see data in Moynier et al., 2015). A total of seven chondrules from Allende were analysed in this study, and these samples were hand-separated using a binocular microscope.

*2.2 Sample purification and calcium separation*

For the bulk meteorites and chondrules, between 0.8 and 20 mg of sample powder was dissolved in a 3:1 mixture of concentrated hydrofluoric acid and nitric acid ($HF/HNO_3$) and placed in small Teflon bombs under pressure on a hot plate (120 °C) for two days. The samples were then dried and dissolved again in 6N HCl to destroy fluoride complexes.

The chemical purification follows a modified method described in Valdes et al. (2014). The samples were loaded in 1 N HNO$_3$ on 1.8 mL of pre-cleaned Eichrom DGA resin to separate Ca and Sr from matrix elements. We found that adding a second elution step using 1% H$_2$O$_2$ in concentrated HNO$_3$ generally decreased the width of the Ca elution peak. Hydrogen peroxide (H$_2$O$_2$) in solutions was used on the DGA resin by Zhang et al. (2011) to elute titanium (Ti) and separate Ti and Zn from the Zr and Hf fraction. These authors loaded 10mL of 12 N HNO$_3$ to elute Ca followed by 10 mL of 12 N HNO$_3$ + 1% H$_2$O$_2$ to collect Ti. We found that adding a step of 3 mL of 15 N HNO$_3$ + 1% H$_2$O$_2$ allowed collection of Ca still attached on the resin. This step includes collection of a minor total fraction of Ti. The inclusion of a minor amount of Ti was deemed acceptable for our measurements, since the interference is on $^{48}$Ca, which was not needed for assessing Ca isotopic compositions between planetary materials. This DGA column step was repeated up to three times to ensure clean separation of Ca from matrix elements. We found that the DGA resin can lose capacity after ~10 uses, so we therefore changed the resin after five passes. After the final DGA column pass, the sample residue was dissolved in 1 mL of concentrated HNO$_3$ to dissolve possible traces of organic matter. Finally, Ca was separated from Sr using 300 μL of Sr-specific resin (Eichrom; 20-50 μm) in Teflon micro-columns following the procedure described in Valdes et al. (2014). One of the difficulties in measuring Ca isotopic compositions is the interference with strontium (Sr). $^{86}$Sr and $^{88}$Sr are the most abundant Sr isotopes and their doubly charged ions interfere with $^{43}$Ca and $^{44}$Ca, respectively. Therefore, the Sr had to be removed totally from the Ca cut. The absence of Sr was checked prior to analysing the samples and is evident from the mass-dependency between the $^{42}$Ca/$^{44}$Ca and $^{43}$Ca/$^{44}$Ca (see below). Collections of

Ca were then dried down and dissolved in 0.1 N HNO$_3$ for mass spectrometric analysis.

*2.3 Mass spectrometry*

The Ca isotopic composition measurements were performed on a *Thermo Fisher Neptune* multi collector inductively coupled plasma (MC-ICP-MS) located at either the Institut de Physique du Globe de Paris, or at Washington University in St Louis following the procedure described in Valdes et al. (2014). The intensities of the $^{42}$Ca, $^{43}$Ca and $^{44}$Ca ion beams were measured. Due to the interference of $^{40}$Ar inherent to traditional MC-ICP-MS, the $^{40}$Ca ion beam was not measured in this study. All the samples were introduced to the plasma of the mass spectrometer through an Apex desolvating introduction system. Measurements were made in medium resolution. Full procedural blanks were below 3 mV on $^{44}$Ca (that is ~3.5ng), which is insignificant relative to the signal of $^{44}$Ca from samples (>2 V).

Standard-sample bracketing was used to correct for instrumental drift over time. The standard used in this study is NIST SRM 915b as the previous international standard used for Ca (NIST SRM 915a) is out of stock. Calcium isotopic variations are defined in per mil units:

$$\delta^{x/y}Ca = \left[ \frac{(\frac{x_{Ca}}{y_{Ca}})_{sample}}{(\frac{x_{Ca}}{y_{Ca}})_{standard}} - 1 \right] \times 1000 \qquad (Eq.\ 1)$$

Where x and y can either be equal to 40, 42, 43, or 44.

To facilitate comparison with literature data, Ca isotopic compositions are presented as $\delta^{44/40}$Ca relative to SRM 915a. After ensuring that all the data were strictly mass-dependent, we converted the results using the mass-dependent fractionation law:

$$\delta^{44/40}Ca = -2 \times \delta^{42/44}Ca \qquad (Eq.\ 2)$$

The Ca isotopic composition was then renormalized in SRM 915a using the relative isotopic composition between SRM 915a and SRM 915b from Valdes et al. (2014) of −0.71‰ for $^{44}Ca/^{40}Ca$. The analytical uncertainty is estimated from repeated measurements of the same solutions and is reported as the 2 standard error (2se), where 2se = 2 × standard deviation/$\sqrt{n}$ for n the number of measurements, and is typically 0.05‰. In addition, the standard reproducibility was tested by bracketing the standard to itself and the variation for $\delta^{42/44}Ca$ and $\delta^{43/44}Ca$ was usually <0.02 ‰. To test accuracy over time, repeated dissolutions of multiple samples were made, processed through the entire chemistry, and measured over multiple analytical sessions. In all cases, replicate dissolutions are isotopically identical within uncertainty.

## 3. Results

Calcium isotopic compositions are presented in Table 1 and Table 2. All samples fall within error of the mass-dependent fractionation line of slope 2.0476 in a $\delta^{42/44}Ca$ versus $\delta^{43/44}Ca$ plot, implying that the isotopic fractionation is mass-dependent (Figure 1). To compare the Ca isotopic composition of samples with literature data, we converted the data to be reported relative to SRM 915a, and the data will only be discussed in terms of $\delta^{44/40}Ca$ normalized to SRM 915a in the following text. For terrestrial basalts, $\delta^{44/40}Ca$ ranges from 0.80‰ to 0.91‰ (Table 1), which is consistent with previous terrestrial basalt data ($\delta^{44/40}Ca$ from 0.75‰ to 1.07‰; Huang et al., 2010; Valdes et al., 2014; Schiller et al., 2015). Terrestrial peridotites are slightly heavier than terrestrial basalts and their $\delta^{44/40}Ca$ vary between 1.08‰ and 1.20‰ (Table 1), which is comparable with the range ($\delta^{44/40}Ca$ = 0.96‰ to 1.15‰) observed in previous studies of mantle peridotites (Amini et al., 2009; Huang et al., 2010; Simon and DePaolo, 2010; Kang et al., 2015). These results are consistent with the data reported for a wide range of basalts and mantle peridotites suggesting a fractionation of Ca isotopes by ~0.1 to 0.3‰ during igneous processes (Amini et al., 2009; Huang et al., 2010; Kang et al. 2015).

Chondrites have Ca isotopic compositions ranging from $\delta^{44/40}$Ca = 0.10‰ to 1.06‰ (Table 1). The data reported for the different groups of carbonaceous chondrites is consistent with previous work (Simon and DePaolo, 2010; Valdes et al., 2014, Schiller et al., 2015; Huang and Jacobsen, 2016, see Figure 2). The isotopic compositions of Allende (CV3) and Indarch (EH4) are consistent with data reported for the same meteorites (Huang and Jacobsen, 2016, Simon and DePaolo, 2010 and Valdes et al., 2014). The three independent dissolution and analysis treatments of Allende show an isotopic variability of ~0.3‰ that is due to sample heterogeneity. The Ca isotopic compositions for the previously unstudied samples Asuka 881595 (CR2), Yamato 980115 (CI1), Khaipur (EL6) and the three CM chondrites fall within the range of each group as defined by Valdes et al. (2014).

All the carbonaceous chondrites studied here (CV, CR, CI and CM) are enriched in the lightest isotopes of Ca relative to the BSE. Enstatite chondrites are within the terrestrial range, similar to data from Huang and Jacobsen (2012) and Valdes et al. (2014). CI and CM chondrites show limited variation within their respective groups.

The $\delta^{44/40}$Ca values of Allende chondrules (CV3) vary between 1.00‰ and 1.21‰ (See Table 2) with an average of 1.10 ± 0.10, which are slightly heavier than the single previous data point from the literature ($\delta^{44/40}$Ca = 1.05‰, Simon et al., 2016). Chondrules are enriched in the heaviest isotopes of Ca relative to bulk chondrites and fall within the range of the estimated BSE. The CAI of Allende is isotopically light ($\delta^{44/40}$Ca = −0.80‰), which is consistent with previous data (Niederer and Papanastassiou, 1984; Huang et al., 2012; Simon et al., 2016).

## 4. Discussion

In the following discussion, our data are compared with the estimated BSE value from Huang et al. (2010). Their estimation is based on the relative proportion of the two major Ca bearing minerals in the mantle peridotites (orthopyroxene and clinopyroxene). Their estimate of the upper mantle ($\delta^{44/40}$Ca = 1.05 ± 0.04‰) is consistent with the most primitive peridotites analysed by Kang et al. (2015) ($\delta^{44/40}$Ca = 1.04 ± 0.12‰) and with our two data for a depleted mantle xenolith sample that lacks clinopyroxene from Lanzarote, Spain (1.08 ± 0.06‰ and 1.17 ± 0.11‰).

The different groups of carbonaceous chondrites have distinct Ca isotopic compositions and, with the exception of CO, are isotopically distinct from the BSE (Figure 2). Since Ca is a lithophile element, the isotopic composition of the BSE corresponds to the isotopic composition of the bulk Earth. Furthermore, since Ca is a refractory major element, it should represent the isotopic composition of the material that accreted to form the Earth. The Ca isotopic difference between the BSE and the CI, CM, CV and CR carbonaceous chondrite groups suggests that these bulk chondrites cannot be considered as major representatives of the condensed matter that accreted to form the Earth. Remarkably, this is only the second example (after O; Clayton and Mayeda, 1999) where a major element shows a large (> 100ppm/amu) isotopic fractionation among chondrite groups and between chondrites and the BSE value. The similarity between the Ca isotopic composition of enstatite chondrites and of the Earth confirms the same observation made by Huang and Jacobsen (2016), Huang and Jacobsen (2012) and Valdes et al. (2014).

Chondrules are enriched in the heavier isotopes of Ca by ~0.7‰ when compared to the bulk of the CV3 chondrite, Allende (0.2‰ < $\delta^{44/40}$Ca < 0.4‰; see Table 1), and the CAIs (−5.60‰ < $\delta^{44/40}$Ca < +0.35‰; Huang et al., 2012).

It is possible to predict the Ca isotopic composition of the matrix by taking the average Ca isotopic compositions for Allende bulk ($\delta^{44/40}$Ca = 0.44‰), chondrules ($\delta^{44/40}$Ca = 1.10‰) and CAIs ($\delta^{44/40}$Ca = −1.25‰). Using Ca concentrations of Allende matrix (17.2 mg/g); chondrules (16.2 mg/g) and CAIs (86.5mg/g) (Grossman and Ganapathy, 1976; Rubin and Wasson, 1987) and the relative proportion of these reservoirs (53% of chondrules, 38% of matrix and 3% of CAIs; Hezel et al., 2008, Ebel et al., 2016), we calculate $\delta^{44/40}$Ca = 0.24‰ for Allende matrix. The matrix of Allende should therefore have a Ca isotopic composition similar to the bulk. The Ca isotopic variation between the different groups of carbonaceous chondrites may be explained in part by the varying proportions of chondrules, which represent an isotopically heavy reservoir of Ca. The heavy isotopic composition of chondrules would also explain the heavier Ca isotopic composition of ordinary chondrites compared to carbonaceous chondrites from the chondrule-poor groups (e.g. CI and CM).

The enrichment of heavy isotopes in some moderately volatile elements (e.g. Zn, Paniello et al., 2012) is usually considered to result from evaporation processes, but Ca, due to its refractory condensation temperature, is not affected by evaporation. This implies that, instead, the nebular gas was already depleted in light calcium isotopes, due to prior condensation of CAIs or other materials through kinetic fractionation effects. If extensive evaporation occurs at relatively high temperature

(2100 K - 2175 K; Richter et al., 2007) calcium would fractionate, creating light isotopes enrichment initially. Following condensation onto preexisting grains, the residual dust would then become enriched in heavy calcium, producing the heavy isotopic composition. However, theses extreme temperatures require complete loss of moderately refractory elements as a collateral effect, as previously pointed out by Simon and DePaolo, (2010). This would only be possible in a nebular shock wave model where the shock front is extremely short-lived (Desch and Connolly, 2002).

The enrichment in lighter Ca isotopes of CAIs has already been observed (Niederer and Papanastassiou, 1984; Huang et al., 2012). This enrichment has also been shown for Ba (Moynier et al., 2015), Sr (Moynier et al., 2010) and Eu (Moynier et al., 2006). Two processes have been proposed to explain this observation: kinetic isotope fractionation due to condensation process (Huang et al., 2012, Simon and DePaolo, 2010) or electromagnetic sorting of isotopes in the nebular gas (Moynier et al., 2006). Moynier et al. (2006) suggested that the distinct isotopic trends for Sm and Eu exclude the possibility of kinetic isotope fractionation and proposed electromagnetic isotope fractionation instead. The latter process relies on elements with low ionization potential (e.g. Ca, Eu, Sr, Ba) that would be preferentially fractionated. This effect is observed for Sm, Eu and Ba elements and for other rare earth elements (REE). Similarly, Simon et al. (2016) measured both Ti and Ca in several CAIs and found that while Ca is highly enriched in the lighter isotopes, Ti is not.

The similarity in Ca isotope composition between chondrules and the Earth has important implications for models that include chondrule-rich precursors for the terrestrial planets (Johansen et al., 2015). Forming terrestrial planets from the accretion of dust into meter-sized and larger objects has long been recognized as problematic due to the high kinetic energies of particle collisions relative to energies of gravitational or chemical attraction (e.g. Blum and Wurm, 2008). More recent models have attempted to overcome these difficulties by making planetary embryos directly from cm-sized objects (after the so-called pebble accretion models originally used to explain the rapid growth of gas giant cores; Lambrechts and Johansen, 2012). Johansen et al. (2015) show that streaming instabilities can concentrate cm-sized particles in high-density regions of the turbulent disk and lead to the formation of ~50 km diameter asteroids and larger embryos the size of Mars in ~3 Ma; a timescale concordant with the estimated accretion time-scale for Mars (Dauphas and Pourmand, 2011). Therefore, chondrule accretion is an appealing mechanism that would enable a rapid growth of planetary embryos in the terrestrial planet formation region. Planets such as Earth would subsequently be formed by collisions between these planetary embryos. In addition to removing a long-standing problem in planetary accretion, making planetary embryos in volatile-poor chondrule-rich regions of the disk can potentially also explain the origin of the volatile depletion observed in terrestrial planets (e.g. Connelly and Bizzarro, 2016).

Since Ca is the only major element that shows large mass-dependent isotopic fractionation between meteorite groups and between meteoritic components (chondrules, matrix and CAIs) it is therefore well suited to test this model. The Ca isotopic composition of chondrules from Allende ($\delta^{44/40}$Ca =1.10 ± 0.10‰) is slightly heavier (while not resolvable within error) than for the BSE ($\delta^{44/40}$Ca =1.05 ± 0.04‰, Huang et al. 2010) and for bulk silicate Mars ($\delta^{44/40}$Ca =1.04 ± 0.09‰, Magna et al. 2015). Based on Ca isotopes this would suggest that the material that formed planetary bodies would permissibly included up to 90% of chondrules, consistent with the growth of embryos reaching the size of Mars after ~4 millions years (Johansen et al., 2015). Due to the variable composition and possible abundance of the matrix, CAIs and chondrules, several mixing scenarios between chondrule-rich planetary embryos and unprocessed outer solar system material (taken to be equivalent to the matrix of the CI chondrites) could be used to calculate the composition of the Earth. For example, mixing 92% of chondrules (45% of Ca with $\delta^{44/40}$Ca =1.10‰) with 8% of matrix (34% of Ca with $\delta^{44/40}$Ca = 0.32‰) would satisfy the Ca mass balance. Therefore, the Ca isotopic composition of meteoritic components demonstrates that accretion of chondrules to form planetary embryos is a permissible mechanism with which to form planets.

A slight difference between the Ca isotopic composition of Mars and the Earth should be observed, as there is ~10% of remaining 'matrix' material that accreted to Earth per this scenario (considering Mars itself as a planetary embryo made with 100% of chondrule-rich materials). We would expect the Ca isotopic composition of Mars to be closer to the isotopic composition of the chondrules than to the isotopic composition of the BSE. Nevertheless, current levels of precision on the estimate of the martian mantle prevents a distinction between the Ca isotopic composition of Earth and Mars since the accretion of 10% of Solar System materials to the Earth might represent only a slight change in isotopic composition. Figure 5 shows a simulation of the Ca isotopic composition relative to different accreted material. This simulation starts with 90% of chondrules accretion followed by progressive addition of 10% of CV, CI, ordinary and enstatite chondrites and differentiated materials (eucrite) accretion. Adding 10% of CV, CI, OC or eucrite reproduce the isotopic composition of the BSE estimate within error.

The model of chondrule-rich planetary embryos would have further collateral isotopic and chemical effects that can be tested. In particular, volatile element abundances vary among terrestrial planets and meteorites (e.g. O'Neill and Palme, 2008). To a first order, the abundance of moderately volatile elements is correlated with their condensation temperature: the CV chondrites are depleted in volatile elements when compared to the CI chondrites, ordinary chondrites are more depleted than CV, and the BSE value is depleted still further. While there are a limited number of elemental abundance data for individual chondrules, and these data are limited to the most abundant volatile elements (e.g. Mn, K, and Zn, see Table S1), it is apparent that chondrules are generally depleted in volatile elements compared to the chondrite matrix (e.g. Bland et al., 2005) and it seems that the volatile element depletion increases with the amount of chondrules in the carbonaceous chondrites (Bland et al., 2005; Zanda et al., 2006). Furthermore, chondrules typically exhibit moderately volatile element abundances that approach the estimate of the composition of the primitive mantle (e.g. Figure 4). Only the siderophile or chalcophile volatile elements (Au, As, Se) show an extreme depletion in the primitive mantle compared to chondrules, consistent with the incorporation of these elements into Earth's core (e.g., Day et al., 2016). Bromine (Br) is the only lithophile highly volatile element for which data are available that shows a strong depletion in Earth compared to chondrules. This could be explained by the loss of volatile elements during the accretion of Earth (O'Neill and Palme, 2008; Paniello et al., 2012; Pringle et al., 2014). Gallium is a siderophile element but for enigmatic reasons behaves as a lithophile element on Earth, plotting directly on the planetary volatility trend (McDonough, 2003). Hence, for the lithophile moderately volatile elements and Ga there is a similarity between chondrules from carbonaceous chondrites and the primitive mantle (Figure 3), the best match being for CV compared to CO and CR.

The Mg/Si ratio is another widely used tracer to distinguish the parent bodies of Earth (Javoy et al., 1995). Figure 4 presents the Mg/Si ratio of the main components (chondrules and matrix) and the bulk of Allende, Ornans (CO3.4), Manych (LL3.4) and Renazzo (CR2). Allende chondrules clearly match the Earth's composition and there is a tendency for the Earth to be more similar to chondrules than to the matrix.

In terms of isotopic effects, Mo, W and Zn are the three elements for which high precision stable isotopic measurements of individual chondrules and matrix samples are available (Budde et al., 2016a; 2016b; Pringle et al., 2017). Allende chondrules show a complementary behaviour in nucleosynthetic anomalies between chondrules and matrix for both Mo and W (Budde et al., 2016a; 2016b). Since the BSE has similar Mo and W isotopic composition as bulk Allende, the isotopic anomalies recorded in Allende chondrules go against the notion of chondrule-rich precursors for terrestrial planets. However, as suggested by previous workers, the complementarity between matrix and chondrules for Mo and W isotopes may reflect the removal or transfer between matrix and chondrules of isotopically anomalous metallic carrier phases of Mo and W. This would therefore modify the isotopic composition of Mo and W (and potentially of other siderophile and chalcophile elements) without affecting the isotopic composition of Ca and other lithophile elements. Furthermore, a fraction of the Mo and W budget of the Earth's mantle may have been brought to Earth during the late accretion (i.e. after core formation; e.g., Day et al., 2016), which would have modified the isotopic composition of Earth's mantle for these elements without modifying the composition of major lithophile elements such as Ca.

In the case of the Zn isotopic composition, Allende chondrules are enriched in light elements and range from $\delta^{66}Zn$ = +0.18‰ to −0.45‰ with a mean of −0.12 ± 0.39‰ compare to the bulk ($\delta^{66}Zn$ = +0.24 ± 0.12‰) (Pringle et al., 2017) and compare well to the estimated BSE ($\delta^{66}Zn$ = +0.28 ± 0.05‰; Chen et al., 2013). However, the authors suggest that the light Zn isotope signature in chondrules is probably due to the segregation of an isotopically heavy sulfide phase during chondrule formation as the sulfide phase is enriched in the heavy isotopes of Zn compare to the bulk. Moreover, Zn is a chalcophile element and has probably been partly removed from the mantle to the core during differentiation of the Earth changing then the signature of the BSE (Mahan et al., 2017).

Other studies for Mg isotopes have shown more isotopic heterogeneities in the compositions of precursors of chondrules, but on average they tend to overlap with the respective bulk silicate Earth (BSE) compositions. For instance, the Mg isotopic compositions of the least altered chondrules from CM2 chondrites have an average $\delta^{26}Mg$ = -0.39 ± 0.30‰ which overlaps with the average composition for whole-rock chondrites of $\delta^{26}Mg$ = -0.31 ± 0.12‰ (Bouvier et al., 2013). Such similarities in Mg isotopic compositions of chondrules were also found for CV meteorites (Olsen et al., 2016). While CV chondrules have variable isotopic compositions of Mg, they have, on average, the composition of the BSE (Olsen et al. 2016). On the other hand, Cr isotopic anomalies (and possibly Ti), reveal a more complex message. Olsen et al. (2016) show that CR chondrules have an average value, which is shifted toward enrichment in $^{54}Cr$ compared to the BSE and therefore a CR chondrule accretion would not match the Cr isotopic composition of the Earth. This caveat suggests that unlike CV chondrules, CR chondrules would certainly not be a good match for the composition of terrestrial planets and that future work should investigate the specific isotopic composition of pools of chondrules from different carbonaceous chondrites.

## 5. Conclusions

We measured the Ca isotopic composition of bulk carbonaceous and enstatite chondrites, terrestrial basalts and peridotites and an individual CAI and chondrules from Allende. These results confirm the Ca isotope variations among carbonaceous chondrites and their enrichment in lighter isotopes compared to the bulk silicate Earth (BSE) value. Conversely, Allende chondrules are isotopically heavier than bulk chondrites suggesting that the relative proportions of chondrules can lead to significant differences in the isotopic compositions of bulk chondrites. Allende chondrules have similar Ca isotopic compositions, within the uncertainties, to bulk silicate Earth and Mars $\delta^{44/40}Ca$ values. This result, in addition to the chemical similarity between chondrules and the BSE, are permissive of chondrule accretion models for planetary embryos and would imply materials with a high chondrule to matrix ratio (~10:1) may have been responsible for the accretion of Earth and Mars. On the other hand, we only have investigated chondrules from CV chondrites and future work should focus on the isotopic composition of chondrules and matrix from other groups of chondrites.

Acknowledgments:

We thank Justin Simon for constructive comments that greatly improved this manuscript. FM acknowledge funding from the European Research Council under the H2020 framework program/ERC grant agreement #637503 (Pristine) and financial support of the UnivEarthS Labex program at Sorbonne Paris Cité (ANR-10-LABX-0023 and ANR-11-IDEX-0005-02), and the ANR through a chaire d'excellence Sorbonne Paris Cité.

References

Amini, M., Eisenhauer, A., Böhm, F., Holmden, C., Kreissig, K., Hauff, F., Jochum, K.P., 2009. Calcium isotopes ($\delta^{44/40}$Ca) in MPI-DING reference glasses, USGS rock powders and various rocks: evidence for Ca isotope fractionation in terrestrial silicates. Geostand. Geoanal. Res. 33, 231–247.

Bland, P.A., Alard, O., Benedix, G.K., Kearsley, A.T., Menzies, O.N., Watt, L.E., Rogers, N.W., 2005. Volatile fractionation in the early solar system and chondrule/matrix complementarity. Proc. Natl. Acad. Sci. USA, 102, 13755-13760.

Bouvier, A., Wadhwa, M., Simon, S.B., Grossman, L., 2013. Magnesium isotopic fractionation in chondrules from the Murchison and Murray CM2 carbonaceous chondrites. Meteorit. Planet. Sci. 48, 339-353.

Blum, J., Wurm, G., 2008. The growth mechanism of macroscopic bodies in protoplanetary disks. Annu. Rev. Astron. Astrophys. 46, 21-56.

Brearley, A. J., Scott, E. R. D., Keil, K., Clayton, R. N., Mayeda, T. K., Boynton, W. V., Hill, D. H., 1989. Chemical, isotopic and mineralogical evidence for the origin of matrix in ordinary chondrites. Geochim. Cosmochim. Acta 53, 2081-2093.

Budde, G., Kleine, T., Kruijer, T.S., Burkhardt, C., Metzler, K., 2016a. Tungsten isotopic constraints on the age and origin of chondrules. Proc. Natl. Acad. Sci. USA, 113, 11, 2886-2891.

Budde, G., Burkhardt, C., Brennecka, G.A., Fischer-Gödde, Kruijer, T.S., Kleine, T., 2016b. Molybdenum isotopic evidence for the origin of chondrules and a distinct genetic heritage of carbonaceous and non-carbonaceous meteorites. Earth Plan. Sci. Lett. 454, 293-303.

Chen, H., Savage, P. S., Teng, F.-Z., Helz, R. T., & Moynier, F., 2013. Zinc isotope fractionation during magmatic differentiation and the isotopic composition of the bulk Earth. Earth Plan. Sci. Lett., 369-370, 34–42.

Clark Jr, R.S., Jarosewich, E., Mason, B., Nelen, J., Gomez, M., Hyde, J.R., 1971. The Allende, Mexico, meteorite shower. Smith. Contr. Earth Sci. 5, 1-53.

Clayton, R.N., Mayeda, T.K., 1999. Oxygen isotope studies of carbonaceous chondrites. Geochim. Cosmochim. Acta 63, 2089-2104.

Connelly, J.N., Bizzarro, M., 2016. Lead isotope evidence for a young formation age of the Earth-Moon system. Earth Plan. Sci. Lett. 452, 36-43.

Dauphas, N., Pourmand, A., 2011. Hf–W–Th evidence for rapid growth of Mars and its status as a planetary embryo. Nature 473, 489–492.

Dauphas, N., Burkhardt, C., Warren, P.H., Fang-Zhen, T., 2014. Geochemical arguments for an Earth-like Moon-forming impactor. Philosophical Transactions of the Royal Society of London A 372, 2024.

Day, J.M.D., Pearson, D.G., Macpherson, C.G., Lowry, D., Carracedo, J.C. 2010. Evidence for distinct proportions of subducted oceanic crust and lithosphere in HIMU-type mantle beneath El Hierro and La Palma, Canary Islands. Geochimica et Cosmochimica Acta. 74:6565-6589.

Day, J.M.D., Brandon A.D., Walker R.J., 2016. Highly Siderophile Elements in Earth, Mars, the Moon, and Asteroids. Reviews in Mineralogy and Geochemistry. 81:161-238.

DePaolo, D.J., 2004. Calcium isotopic variations produced by biological, kinetic, radiogenic and nucleosynthetic processes. Reviews in Mineralogy and Geochemistry 55, 255-288.

Desch, S.J., and Connolly, H.C. Jr., 2002. A model of the thermal processing of particles in solar nebula shocks: Application to the cooling rates of chondrules. Meteor. Plan. Sci. 37: 183–207.

Dodd, R. T., 1978. The composition and origin of large microporphyritic chondrules in the manych (L-3) chondrite. Earth Plan. Sci. Lett. 39, 52-66.

Ebel, D. S., Weisberg, M. K., Hertz, J., Campbell, A. J., 2008. Shape, metal abundance, chemistry, and origin of chondrules in the Renazzo (CR) chondrite. Meteoritics & Planetary Science 43, 1725-1740.

Ebel, D.s., Brunner, C., Konrad, K., Leftwich, K., Erb, I., Lu, M., Rodriguez, H., Crapster-Pregont, E.J., Friedrich, J.M., 2016. Abundance, major element composition and size of components and matrix in CV, CO and Acfer 094 chondrites. Geochim. Cosmochim. Acta 172, 322-356.

Georg, R.B., Halliday, A.N., Schauble, E.A., Reynolds, B.C., 2007. Silicon in the Earth's core. Nature 447, 1102–1106.

Grossman, L., Ganapathy, R., 1976. Trace elements in the Allende meteorite-I. Coarse-grained, Ca-rich inclusions. Geochim. Cosmochim. Acta 40, 331-344.

Hezel, D.C., Russell, S.S., Ross, A.J., Kearsley, A.T., 2008. Modal abundances of CAIs: Implications for bulk chondrite element abundances and fractionations. Meteorit. Planet. Sci. 43, 1879-1894.

Hezel, D.C., Palme, H., 2010. The chemical relationship between chondrules and matrix and the chondrule matrix complementarity. Earth Planet. Sci. Lett., 294, 85-93.

Huang, S., Farkaš, J., Jacobsen, S.B., 2010. Calcium isotopic fractionation between clinopyroxene and orthopyroxene from mantle peridotites. Earth Planet. Sci. Lett. 292, 337–344.

Huang, S., Jacobsen, S.B., 2012. Calcium isotopic variations in chondrites: implications for planetary isotope compositions. In: 43rd Annual Lunar Planet. Sci. Conference. Abstract #1334.

Huang, S., Farkaš, J., Yu, G., Petaev, M.I., Jacobsen, S.B., 2012. Calcium isotopic ratios and rare earth element abundances in refractory inclusions from the Allende CV3 chondrite. Geochim. Cosmochim. Acta 77, 252–265.

Huang, S., Jacobsen, S., B., 2016. Calcium isotopic compositions of chondrites. Geochim. Cosmsochim. Acta *In Press*. DOI: 10.1016/j.gca.2016.09.039.

Jarosewich, E., Dodd, R.T., 1981. Chemical variations among L-group chondrites, II. Chemical distinctions between L3 and LL3 chondrites. Meteoritics 16, 1, 83-91.

Javoy, M., 1995. The integral enstatite chondrite model of the earth. Geophys. Res. Lett. 22, 2219–2222.

Johansen, A., Mac Low, M., Lacerda, P., Bizzarro, M., 2015. Growth of asteroids, planetary embryos, and Kuiper belt objects by chondrule accretion. Science Advances Vol 1, No. 3.

Kang, J.-T., Zhu, H.-L., Liu, Y.-F., Liu, F., Wu, F., Hao, Y.-T., Zhi, X.-C., Zhang, Z.-F., Huang, F., 2015. Calcium isotopic composition of mantle xenoliths and minerals from Eastern China. Geochim. Cosmochim. Acta 174, 335-344.

Kong, P., Palme, H., 1999. Compositional and genetic relationship between chondrules, chondrule rims, metal, and matrix in the Renazzo chondrite. Geochim. Cosmochim. Acta 63, 3673-3682.

Lambrechts, M., Johansen, A., 2012. Rapid growth of gas giant cores by pebble accretion. Astron. Astrop. 544, A32, 1-13.

Lodders, K., 2003. Solar system abundances and condensation temperatures of the elements. Astrophys. J. 591, 1220–1247.

Magna, T., Gussone, N., Mezger, K., 2015. The calcium isotope systematics of Mars. Earth Plan. Sci. Lett. 430, 86-94.

Mahan, B., Siebert, J., Pringle, E.A., Moynier, F., 2017. Elemental partitioning and isotopic fractionation of Zn between metal and silicate and geochemical estimation of the S content of the Earth's core. Geochim. Cosmochim. Acta 196, 252-270.

McDonough, W.F., 2003. Compositional model for the Earth's core. Treatise on Geochemistry, 2, 47–568.

Moynier, F., Bouvier, A., Blichert-Toft, J., Telouk, P., Gasperini, D., Albarède, F.,

2006. Europium isotopic variations in Allende CAIs and the nature of mass-dependent fractionation in the solar nebula. Geochim. Cosmochim. Acta 70, 4287-4294.

Moynier, F., Agranier, A., Hezel, D., Bouvier, A., 2010. Sr stable isotope composition of Earth, the Moon, Mars, Vesta and meteorites. Earth Planet. Sci. Lett. 300, 359–366.

Moynier, F., Fegley Jr, B., 2015. The Earth's building blocks in Badro, J., Walter, M. J., (Eds), The Early Earth: accretion and differenciation. John Wiley & Sons, Inc., Washington, D.C., pp. 27-47.

Moynier, F., Pringle, E. A., Bouvier, A., Moureau, J., 2015. Barium stable isotope composition of the Earth, meteorites, and calcium-aluminium-rich inclusions. Chem. Geol, 413, 1-6.

Niederer, F.R., Papanastassiou, D.A., 1984. Ca isotopes in refractory inclusions. Geochim. Cosmochim. Acta 48, 1279–1293.

Olsen, M.B., Wielandt, D., Schiller, M., Van Kooten, E.M.M.E., Bizzarro, M., 2016. Magnesium and 54Cr isotope compositions of carbonaceous chondrite chondrules – Insights into early disk processes. Geochim. Cosmochim. Acta 191, 118-138.

O'Neill, H.St.C., Palme, H., 2008. Collisional erosion and non-chondritic composition of the terrestrial planets. Philosophical transactions of the royal society of London 366, 4205-4238.

Palme, H., O'Neill, H. St. C., 2003. Cosmochemical estimates of mantle composition. Treatise on Geochemistry 2, 1-38.

Palme, H., Lodders, K., Jones, A., 2014. Solar system abundances of the elements, Treatise on Geochemistry 2, 15-36.

Paniello, R.C., Day, J.M.D., Moynier, F., 2012. Zinc isotopic evidence for the origin of the Moon. Nature 490(7420), 376–379.

Pringle, E.A., Moynier, F., Savage, P.S., Badro, J., Barrat, J-A., 2014. Silicon isotopes in angrites and volatile loss in planetesimals. Proc. Natl. Acad. Sci. USA 111, 17029-17032.

Pringle, E.A., Moynier, F., Beck, P., Paniello, R., Hezel, D.C., 2017. The origin of volatile element depletion in early solar system material: clues from Zn isotopes in chondrules. Earth Plan. Sci. Lett. 468, 62-71.

Richter, F.M., Janney, P.E., Mendybaev, R.A., Davis, A.M., Wadhwa, M., 2007. Elemental and isotopic fractionation of Type BCAI-like liquids by evaporation. Geochim. Cosmochim. Acta 71 (22), 5544–5564.

Rubin, A. E., Wasson, J. T., 1987. Chondrules, matrix and coarse-grained chondrule rims in the Allende meteorite: Origin, interrelationships and possible precursor components. Geochim. Cosmochim. Acta 51, 1923-1937.

Rubin, A. E., Wasson, J. T., 1988. Chondrules and matrix in the Ornans CO3 meteorite: Possible precursor components. Geochim. Cosmochim. Acta 52, 425-432.

Rubin, A.E., 2011. Origin of the differences in refractory-lithophile-element abundances among chondrite groups. Icarus 213, 547-558.

Russel, W. A., Papanastassiou, D. A., Tombrello, T. A., 1978. Ca isotope fractionation on the Earth and other solar system materials. Geochim. Cosmochim. Acta 42, 1075-1090.

Savage, P.S., Moynier, F., 2013. Silicon isotopic variation in enstatite meteorites: clues to their origin and Earth-forming material. Earth Planet. Sci. Lett. 361, 487–496.

Schiller, M., Paton, C., Bizzarro, M., 2015. Evidence for nucleosynthetic enrichment of the protosolar molecular cloud core by multiple surpernova events. Geochim. Cosmochim. Acta 149, 88-102.

Simon, J.I., DePaolo, D.J., 2010. Stable calcium isotopic composition of meteorites and rocky planets. Earth Planet. Sci. Lett. 289, 457–466.

Simon, J.I, Jordan M.K., Tappa, M.J., Kohl, I.E., Young, E.D., 2016. Calcium and titanium isotope fractionation in cais : Tracers of condensation and inheritance in the early solar protoplanetary disk. In: 47rd Annual Lunar Planet. Sci. Conference. Abstract #1397.

Teng, F.-Z., Li, W.-Y., Ke, S., Marty, B., Dauphas, N., Huang, S., Wu, F.-Y., Pourmand, A., 2010. Magnesium isotopic composition of the Earth and chondrites. Geochim. Cosmochim. Acta 74, 4150–4166.

Valdes, M. C., Moreira, M., Foriel, J., Moynier, F., 2014. The nature of Earth's building blocks as revealed by calcium isotopes. Earth Planet. Sci. Lett. 394, 135-145.

Weisberg, M. K., Prinz, M., Clayton, R. N., Mayeda, T. K., 1993. The CR (renazzo type) carbonaceous chondrite group and its implications. Geochim. Cosmochim. Acta 57, 1567-1586.

Zanda, B., Hewins, R.H., Bourot-Denise, M., Bland, P.A., Albarède, F., 2006. Formation of solar nebula reservoirs by mixing chondritic components. Earth Planet. Sci. Lett. 248, 650–660.

Zhang, J., Dauphas, N., Davis, A., M., Pourmand, A., 2011. A new method for MC-IPCMS measurements of titanium isotopic composition: Identification of correlated isotope anomalies in meteorites. J. Anal. At. Spectrom., 26, 2197-2205.

Figures captions

Figure 1: Three isotope plot of $\delta^{42/44}Ca$ versus $\delta^{43/44}Ca$ for terrestrial and extraterrestrial materials relative to SRM 915b. Calcium isotopic compositions of chondrites are close to terrestrial basalts. All samples follow a mass dependent equilibrium line (slope = 2.0476).

Figure 2: Calcium isotopic composition of chondrites compared to BSE (represented by the shaded gray area; Huang et al., 2010). Enstatite chondrites are within the BSE range. Carbonaceous chondrites show a variation of Ca isotopic compositions. Allende chondrules are similar to Earth. Data from Valdes et al., (2014) (V14), Simon and DePaolo, (2010) (SD10), Huang and Jacobsen, (2016) (H16), Kang et al., (2015)

(K15), Amini et al., (2009) (A9), Simon et al., (2016) (S16) and Schiller et al., (2015) (S15).

Figure 3: Abundances relative to Mg of the moderately volatile elements Mn, K, Ga, and Zn in components of representative carbonaceous chondrites for the CV (Allende; Figure 3a), CO (Ornans; Figure 3b) and CR (Renazzo; Figure 3c) groups compared to the primitive mantle of Earth (Palme and O'Neill, 2003). Primitive mantle is more similar to the chondrules than matrix. Allende, Ornans and Renazzo data are from Rubin and Wasson, (1987), Rubin and Wasson (1988) and Kong and Palme, (1999), respectively.

Figure 4: Mg/Si ratio of the primitive mantle and of the main components (chondrules and matrix) and bulk of Allende (CV3), Ornans (CO3), Manych (LL3.4) and Renazzo (CR2). The Mg/Si ratio of the primitive mantle is more similar with the chondrules than the bulk and is distinct from the matrix. Allende data from Clark et al., (1971); Rubin and Wasson, (1987) and Hezel and Palme, (2010); Ornans data from Rubin and Wasson, (1988); Manych data from Brearly et al., (1989); Dodd, (1978) and Jarosewich and Dodd, (1981), Renazzo data from Weisberg et al., (1993) and Ebel et al., (2008).

Figure 5: Simulation of the Ca isotopic composition relative to the percentage of accreted material starting with 90% of chondrules. Adding 10% of CV, CI, OC or eucrite material allows our simulation to reach the isotopic composition of the BSE at 100% accreted material. CV are represented by the average of Allende (Valdes et al., 2014 and this study), CI by Orgueil (Valdes et al., 2014 and this study), OC are averaged from LL type (Valdes et al., 2014 and Simon and DePaolo, 2010), eucrite materials is represented by the Juvinas composition (Simon and DePaolo, 2010) and enstatite chondrites are averaged from Valdes et al. (2014) and our data.

Figures

Figure 1

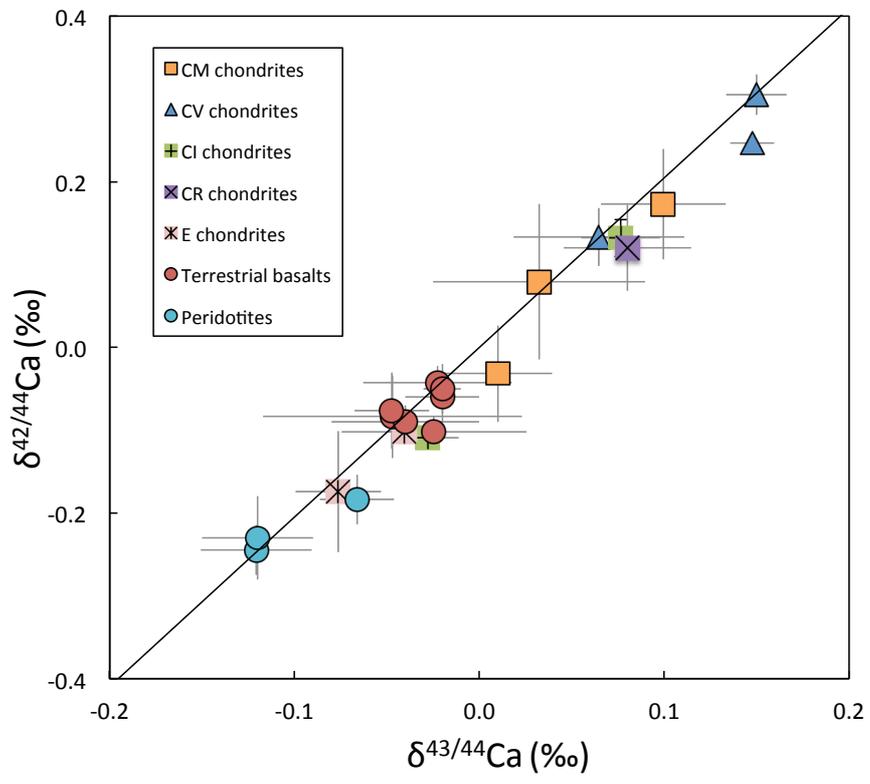

Figure 2

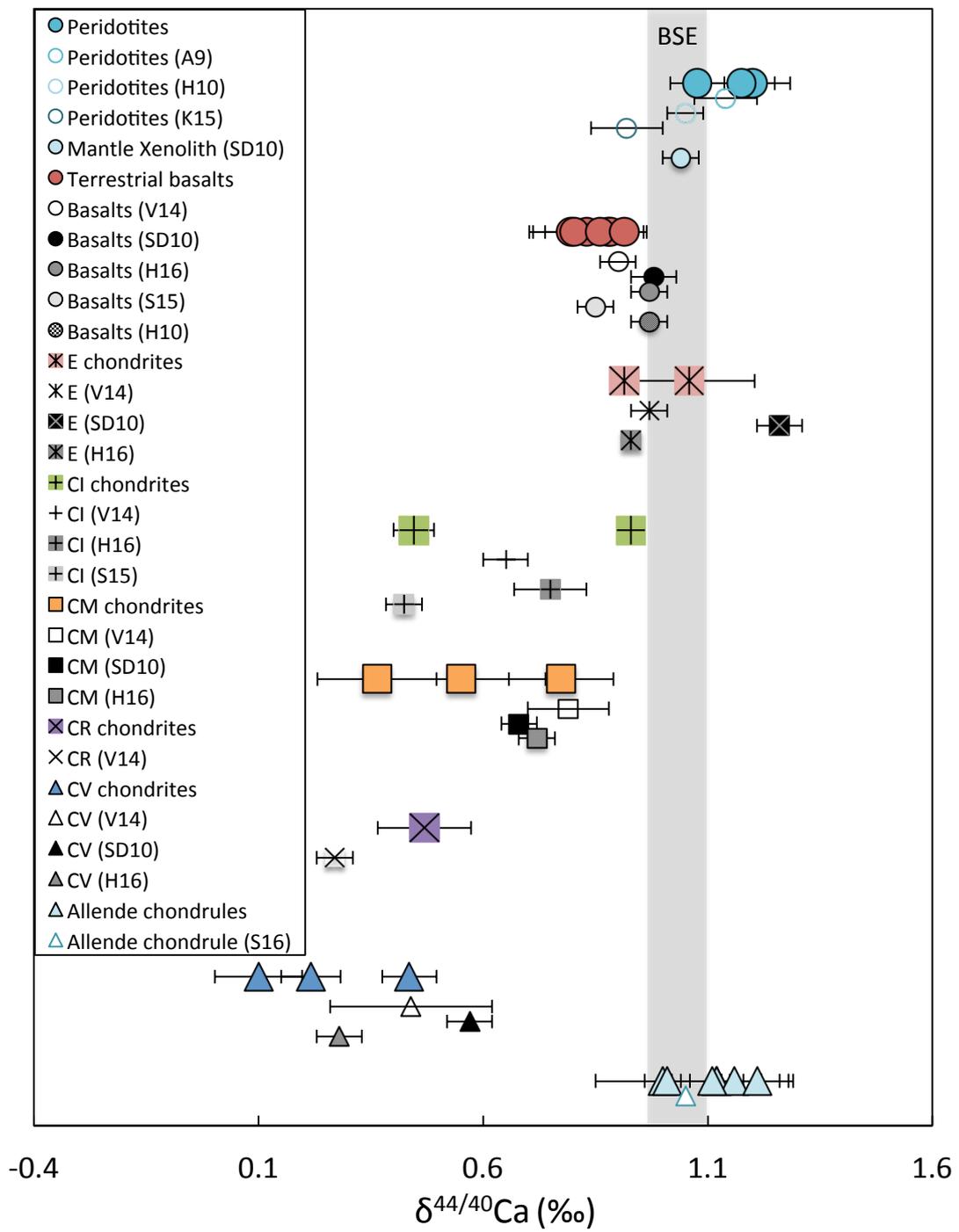

Figure 3

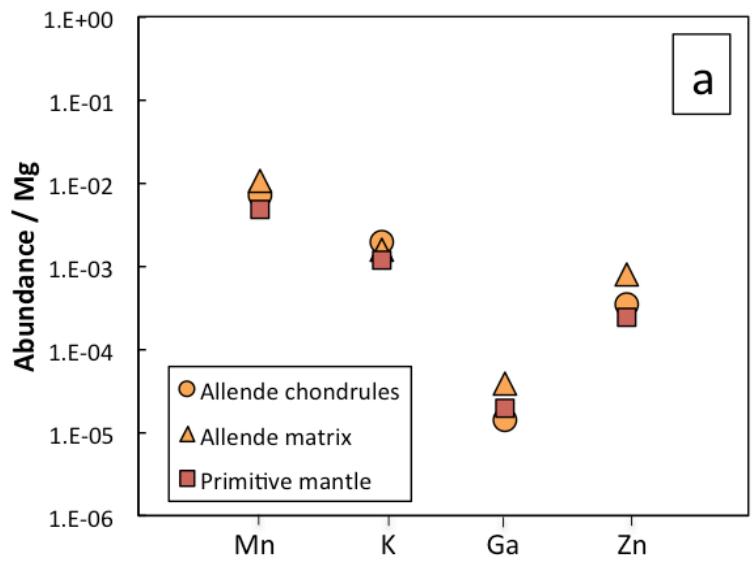
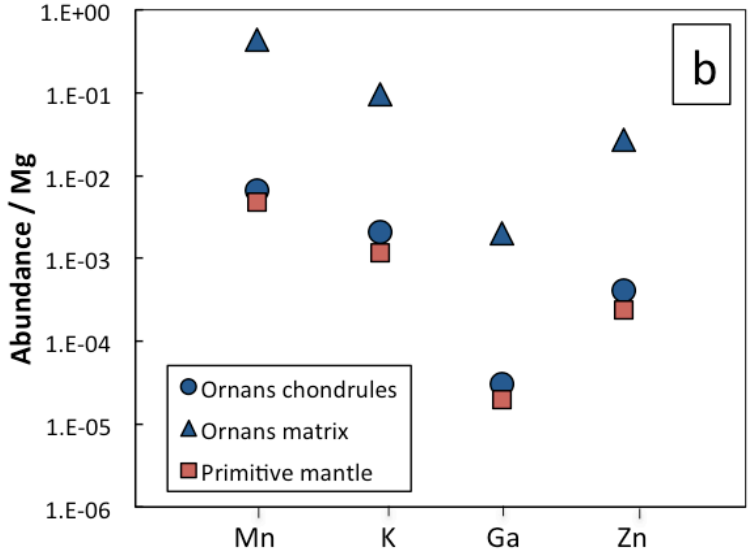
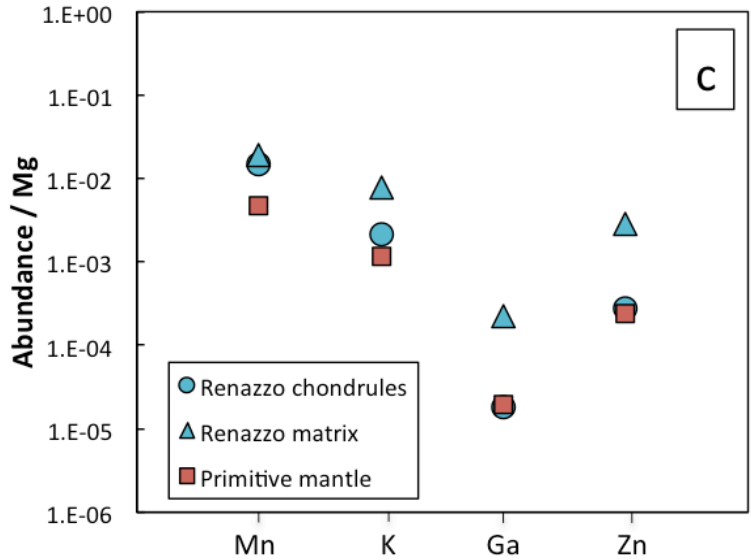

Figure 4

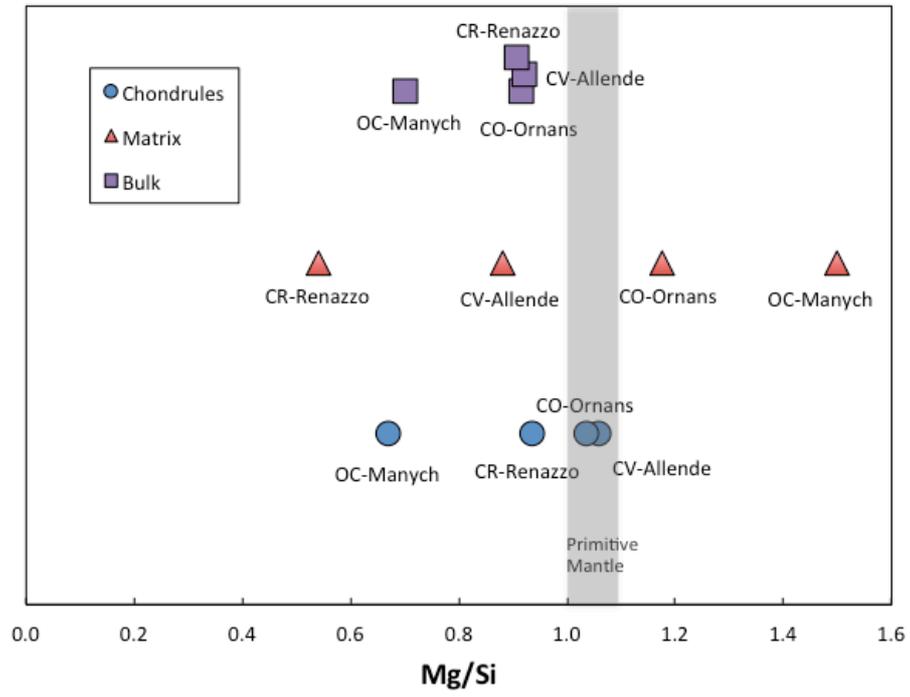

Figure 5

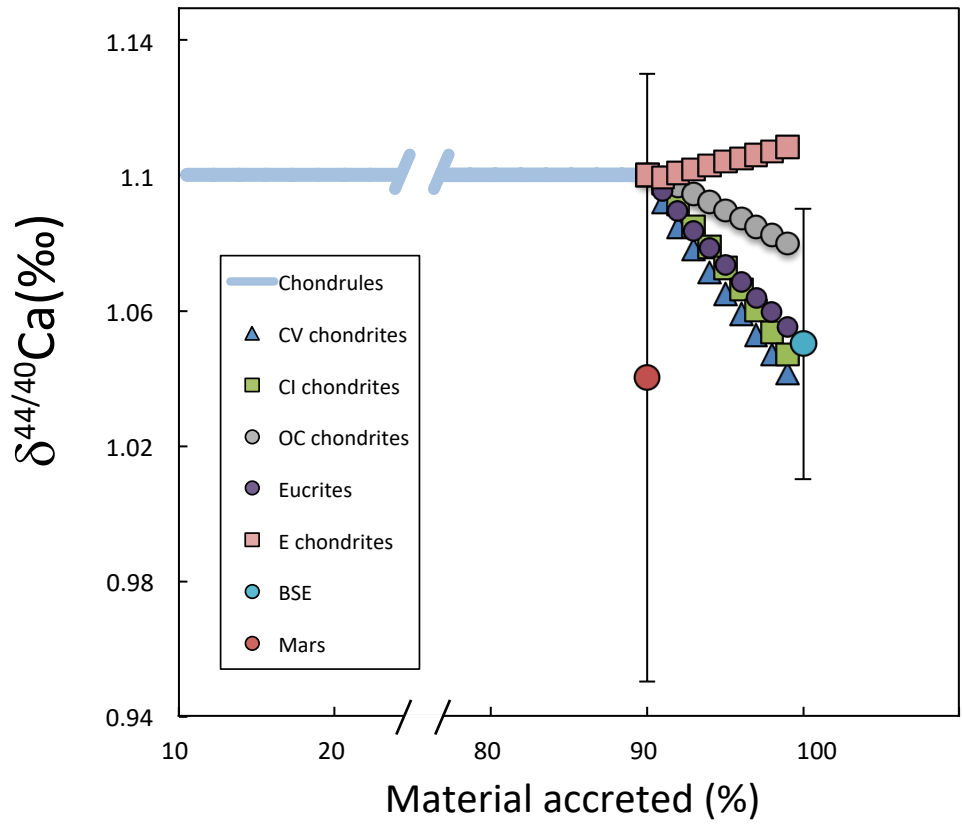

**Table 1 - Calcium isotopic composition of terrestrial basalts, peridotites and carbonaceous and enstatite chondrites relative to SRM915b and renormalized to SRM915a.**

| Sample | Location | $\delta^{42/44}Ca$ [2] | 2se[1] | $\delta^{43/44}Ca$ [2] | 2se[1] | $\delta^{44/40}Ca$ [2] | $\delta^{44/40}Ca$ [3] | 2se[1] | n[4] |
|---|---|---|---|---|---|---|---|---|---|
| **Terrestrial Basalts** | | | | | | | | | |
| EH12 | El Hierro | -0.08 | 0.05 | -0.05 | 0.07 | 0.17 | 0.88 | 0.11 | 6 |
| EH12 replicate | El Hierro | -0.09 | 0.02 | -0.04 | 0.04 | 0.17 | 0.88 | 0.04 | 5 |
| EH12 replicate | El Hierro | -0.06 | 0.03 | -0.02 | 0.02 | 0.12 | 0.83 | 0.05 | 5 |
| EH15 | El Hierro | -0.04 | 0.02 | -0.02 | 0.04 | 0.09 | 0.80 | 0.02 | 6 |
| BHVO | Central Pacific | -0.05 | 0.03 | -0.02 | 0.01 | 0.09 | 0.80 | 0.05 | 6 |
| BCR-2 | Continental US | -0.10 | 0.02 | -0.02 | 0.05 | 0.20 | 0.91 | 0.04 | 5 |
| EW 9309 10D | South Atlantic | -0.08 | 0.05 | -0.05 | 0.02 | 0.15 | 0.86 | 0.09 | 3 |
| *Average* | | *-0.07* | | *-0.03* | | *0.14* | *0.85* | | |
| **Peridotites** | | | | | | | | | |
| PCC | Continental US | -0.24 | 0.03 | -0.12 | 0.03 | 0.49 | 1.20 | 0.05 | 5 |
| LZ0604B | Lanzarote | -0.18 | 0.03 | -0.07 | 0.02 | 0.37 | 1.08 | 0.06 | 4 |
| LZ0604B replicate | Lanzarote | -0.23 | 0.05 | -0.12 | 0.03 | 0.46 | 1.17 | 0.11 | 5 |
| *Average* | | *-0.22* | | *-0.10* | | *0.44* | *1.15* | | |
| **Carbonaceous chondrites** | | | | | | | | | |
| Allende | CV3 | 0.25 | 0.01 | 0.15 | 0.01 | -0.49 | 0.26 | 0.07 | 6 |
| Allende replicate | CV3 | 0.31 | 0.02 | 0.15 | 0.02 | -0.61 | 0.10 | 0.05 | 4 |
| Allende replicate | CV3 | 0.13 | 0.03 | 0.06 | 0.05 | -0.27 | 0.44 | 0.08 | 3 |
| *Average CV* | | *0.23* | | *0.12* | | *-0.46* | *0.27* | | |
| Asuka 881595 | CR2 | 0.12 | 0.05 | 0.08 | 0.03 | -0.24 | 0.47 | 0.10 | 3 |
| PCA 02012.23 | CM | 0.08 | 0.09 | 0.03 | 0.06 | -0.16 | 0.55 | 0.19 | 4 |
| Cold Bokkeveld | CM2 | 0.17 | 0.07 | 0.10 | 0.03 | -0.35 | 0.36 | 0.13 | 3 |
| PCA 02010.06 | CM | -0.03 | 0.06 | 0.01 | 0.03 | 0.06 | 0.77 | 0.12 | 3 |
| *Average CM* | | *0.07* | | *0.05* | | *-0.15* | *0.56* | | |
| Orgueil | CI1 | 0.13 | 0.02 | 0.08 | 0.02 | -0.26 | 0.45 | 0.04 | 4 |
| Yamato 980115 | CI1 | -0.11 | 0.01 | -0.03 | 0.02 | 0.22 | 0.93 | 0.02 | 4 |
| *Average CI* | | *0.01* | | *0.02* | | *-0.02* | *0.69* | | |
| **Enstatite chondrites** | | | | | | | | | |
| Indarch | EH4 | -0.17 | 0.07 | -0.08 | 0.02 | 0.35 | 1.06 | 0.15 | 3 |
| Khairpur | EL6 | -0.10 | 0.02 | -0.04 | 0.01 | 0.20 | 0.91 | 0.03 | 4 |
| *Average E* | | *-0.14* | | *-0.06* | | *0.28* | *0.99* | | |

[1] 2se= 2 x standard deviation /√n
[2] The Ca isotope ratios are normalized relative to SRM 915b
[3] The Ca isotope ratios are normalized relative to SRM 915a
[4] n= number of measurements

Table 2 - Calcium isotopic composition of Allende chondrules and CAI relative to SRM915b and renormalized to SRM915a.

| Sample | Type | $\delta^{42/44}Ca$ [2] | 2se[1] | $\delta^{43/44}Ca$ [2] | 2se[1] | $\delta^{44/40}Ca$ [2] | $\delta^{44/40}Ca$ [3] | 2se[1] | n[4] |
|---|---|---|---|---|---|---|---|---|---|
| **Allende chondrules** | | | | | | | | | |
| EA1 | CV3 | -0.21 | 0.05 | -0.08 | 0.01 | 0.41 | 1.12 | 0.10 | 5 |
| EA2 | CV3 | -0.20 | 0.03 | -0.10 | 0.01 | 0.41 | 1.12 | 0.06 | 5 |
| EA3 | CV3 | -0.22 | | -0.04 | | 0.45 | 1.16 | | 1 |
| EA4 | CV3 | -0.25 | 0.06 | -0.12 | 0.01 | 0.5 | 1.21 | 0.12 | 3 |
| EA5 | CV3 | -0.20 | 0.08 | -0.09 | 0.01 | 0.4 | 1.11 | 0.08 | 3 |
| EA6 | CV3 | -0.14 | 0.07 | -0.06 | 0.04 | 0.29 | 1.00 | 0.15 | 3 |
| EA7 | CV3 | -0.15 | 0.08 | -0.03 | 0.03 | 0.3 | 1.01 | 0.15 | 3 |
| *Average* | | *-0.21* | | *-0.07* | | *0.42* | *1.10* | | |
| **CAI Allende** | | | | | | | | | |
| AB1 | CV3 | 0.76 | 0.04 | 0.38 | 0.03 | -1.51 | -0.80 | 0.08 | 5 |

[1] 2se= 2 x standard deviation /√n
[2] The Ca isotope ratios are normalized relative to SRM 915b
[3] The Ca isotope ratios are normalized relative to SRM 915a
[4] n= number of measurements

Table S1 - Concentration in ppm of moderately volatile elements in the primitive mantle and in the main components of Allende (CV3), Ornans (CO3) and Renazzo (CR2). Primitive mantle data are from Palme and O'Neill 2003. Allende, Ornans, Renazzo data are from respectively Rubin and Wasson, 1987, Rubin and Wasson 1988 and Kong and Palme, 1999.

|  | Mn | As | Au | K | Ga | Zn | Se | Br |
|---|---|---|---|---|---|---|---|---|
| **Primitive Mantle** | 1050 | 0.066 | 0.00088 | 260 | 4.4 | 53.5 | 0.079 | 0.075 |
| **Allende** | | | | | | | | |
| Chondrules | 1449 | 0.83 | 0.09 | 409 | 2.9 | 71.2 | 9.76 | 1.93 |
| Matrix | 1655 | 1.95 | 0.16 | 252 | 6.1 | 125 | 8 | 1.95 |
| **Ornans** | | | | | | | | |
| Chondrules | 1329 | 1.8 | 0.19 | 418 | 6.06 | 81.5 | 18.42 | 3.88 |
| Matrix | 1810 | 2.65 | 0.27 | 400 | 8.5 | 115 | 9.6 | 1.7 |
| **Renazzo** | | | | | | | | |
| Chondrules | 2300 | 1.19 | 0.12 | 338 | 2.9 | 44.1 | 5.54 | 0.95 |
| Matrix | 1690 | 1.48 | 0.14 | 668 | 19.5 | 249 | 15.2 | 1.75 |